\documentclass[11pt]{revtex4}
\usepackage{amssymb,epsf}                      %revised 21 Feb 2013%
\usepackage{latexsym}
\begin{document}
\title{Extremal Myers-Perry black holes in
Born-Infeld-dilaton theory}

\author{Masoud Allaverdizadeh$^{1}$\footnote{masoud.alahverdi@gmail.com},
Seyed H. Hendi $^{2,3}$\footnote{hendi@shirazu.ac.ir},  Ahmad
Sheykhi$^{2,3}$\footnote{asheykhi@shirazu.ac.ir}}
\address{ $^1$  Centro Multidisciplinar de Astrof\'{\i}sica -
CENTRA, Departamento de F\'{\i}sica,
Instituto Superior T\'ecnico - IST, Universidade T\'ecnica de Lisboa -
Av. Rovisco Pais 1, 1049-001 Lisboa, Portugal. \\
          $^2$  Physics Department and Biruni Observatory,
College of Sciences, Shiraz University, Shiraz 71454, Iran\\
          $^3$ Center for Excellence in Astronomy and 
Astrophysics (CEAA-RIAAM), Maragha P. O. Box 55134-441, Iran}

\begin{abstract}
\vspace*{1.5cm} \centerline{\bf Abstract} \vspace*{1cm}

A class of perturbative charged rotating dilaton black hole
solutions coupled to the nonlinear Born-Infeld electrodynamics is
presented. The strategy for obtaining these solutions is to start
from Myers-Perry solutions, and then considering the effects of the
dilaton field, Born-Infeld parameter as well as the electric charge as a perturbative
parameter. We perform the perturbations up to 4th order in the extremal black holes of Einstein-Born-Infeld-dilaton gravity with equal angular momentum and arbitrary dilaton coupling constant. We discuss the physical properties of these black
holes and study their dependence on the perturbative parameter
$q$, the dilaton coupling constant $\alpha$ and the Born-Infeld
parameter $\beta$. One can see these black holes solutions satisfy a generalized
Smarr relation.
\end{abstract}
 \maketitle
\newpage
\section{Introduction\label{Intro}}
A scalar field called the dilaton appears in the low energy limit
of string theory. The presence of the dilaton field has important
consequences on the causal structure, asymptotic behavior and thermodynamic properties of the black holes. Thus much interest has been focused on the
study of the dilaton black holes in recent years. Exact solutions
for charged dilaton black holes in which the dilaton is coupled to
the Maxwell field have been constructed by many authors. The
asymptotically flat Einstein-Maxwell-dilaton (EMD) solutions  with
spherical horizon have been investigated in \cite{Gibbons}. They
have shown that the physical properties of the black holes depend
sensitively on the value of the dilaton coupling constant.
Solutions of EMD theory with one Liouville-type potential which
are neither asymptotically flat nor anti-de Sitter (AdS) have been
studied in \cite{Chan,Shey0}. It was shown that with  an
appropriate combination of three Liouville-type potentials one can
derive exact solutions in EMD theory \cite{Gao,Shey1}. In addition, investigation of a class of higher dimensional topological charged dilatonic black holes in the background of the (AdS) universe and their thermodynamic properties has been presented in \cite{EMDads}.

Besides, recently more interest has been focused on the gravity
theories with Born-Infeld (BI) action \cite{Born-Infeld} due to
its several remarkable features
\cite{Wiltshire1,Fradkin1,Tseytlin1,Cataldo1,Gibbons1}. For
example, BI type effective actions, which arise naturally in open
super strings and D branes are free from physical singularities.
Exact solutions for charged static BI black holes have been
considered by many authors \cite{Sheykhi1}. However, finding
rotating BI black holes solutions in higher dimensions is a
difficult task due to the size and complexity of the equations.
For these reasons the number of rotating BI solutions known
analytically form is very limited \cite{BIrot,BIdrot} and as far as authors know general solutions of charged rotating EBId black holes have not been studied so far. This forces us to use
alternative technique to study these black holes. Employing
higher order perturbation theory, odd dimensional charged rotating
black holes in Einstein-Maxwell \cite{Allahverdi1} and
Einstein-Maxwell-dilaton theory \cite{Allahverdi2} have been
studied. Recently, we have used the perturbative method to study
the extremal Einstein BI black holes \cite{Allahverdi3}. In this
paper, we extend our previous study to dilaton gravity.  Following
the approach we developed in \cite{Allahverdi3}, we find charged
rotating extremal BI black holes solutions coupled to a  scalar
dilaton field. Again, we focus on black holes at extremality.
Starting from the Myers-Perry black holes, we evaluate the
perturbative series up to $4$th order in the charge for black
holes in 5 dimensions. We determine the physical properties of
these black holes for general dilaton coupling constant. In
particular, we investigate the effects of the presence of the
dilaton field, BI parameter, and the perturbative charge
parameter, on the gyromagnetic ratio of these rotating black
holes.

The remainder of this paper is outlined as follows. In section
\ref{FIRST}, we present the basic field equations of nonlinear BI
theory in dilaton gravity and obtain a new class of perturbative
charged rotating solutions in five dimensions. In section
\ref{PQ}, we calculate the physical quantities of the solutions
such as mass, angular momentum, dilaton charge, electric charge
and the gyromagnetic ratio. The last section is devoted to summary
and conclusions.
\section{Metric and Gauge Potential\label{FIRST}}
We examine the 5-dimensional action in which gravity is coupled to
dilaton and BI fields
\begin{eqnarray}
S &=&\frac{1}{16 \pi}\int d^{5}x\sqrt{-g}\left( R-\frac{4}{3}(\nabla\Phi)^2\text{
}+L(F,\Phi)\right),  \label{Lag}
\end{eqnarray}
where ${R}$ is scalar curvature, $\Phi$ is the dilaton field and the last term $L(F,\Phi)$ of the action is given by
\begin{eqnarray}
L(F,\Phi)
&=&4\beta^{2}e^{(\frac{4\alpha\Phi}{3})}\left(1-\sqrt{1+\frac{e^{(\frac{-8\alpha\Phi}{3})}F^{\mu
\nu }F_{\mu \nu }}{2 \beta^{2}}}\right). \label{BI}
\end{eqnarray}
Here, $\alpha$ is a constant determining the strength of coupling of the scalar and electromagnetic fields, where $F_{\mu \nu }=\partial _{\mu }A_{\nu }-\partial _{\nu
}A_{\mu }$ is the electromagnetic field tensor,  $A_{\mu }$ is
the electromagnetic vector potential, and $\beta $ is the BI parameter
with units of mass. In the limit $\beta \rightarrow \infty
$, $L(F,\Phi)$ reduces to the Lagrangian of
the standard Maxwell field coupled to a dilaton field
\begin{eqnarray}
L(F,\Phi) &=&-e^{(\frac{-4\alpha\Phi}{3})}F^{\mu \nu }F_{\mu \nu
}, \label{MD}
\end{eqnarray}
and on the other hand, $L(F,\Phi) \rightarrow 0$ as $\beta \rightarrow 0$. It is convenient to set
\begin{eqnarray}
L(F,\Phi) &=&4\beta^{2}e^{(\frac{4\alpha\Phi}{3})}L(Y), \label{BI1}
\end{eqnarray}
where
\begin{eqnarray}
L(Y) &=&1-\sqrt{1+Y}, \label{L}
\end{eqnarray}
\begin{eqnarray}
Y &=&\frac{e^{\frac{-8\alpha\Phi}{3}}F^{\mu \nu }F_{\mu \nu
}}{2\beta^2}. \label{Y}
\end{eqnarray}
The equations of motion can be obtained by varying the action with
respect to the gravitational field $g_{\mu \nu }$, the dilaton
field $\Phi$ and the gauge field $A_{\mu }$ which yield
the following field equations
\begin{equation}
R_{\mu \nu }=
\frac{4}{3}\partial_{\mu}\Phi\partial_{\nu}\Phi-4e^{\frac{-4\alpha\Phi}{3}}\partial_{Y}L(Y)F_{\mu \eta }F_{\nu }^{\text{
}\eta }+\frac{4\beta^2}{3}e^{\frac{4\alpha\Phi}{3}}[2Y\partial_{Y}L(Y)-L(Y)]g_{\mu \nu } ,
\label{FE1}
\end{equation}
\begin{equation}
\nabla^{2}\Phi=
2\alpha\beta^2e^{\frac{4\alpha\Phi}{3}}[2Y\partial_{Y}L(Y)-L(Y)], \label{FE2}
\end{equation}
\begin{equation}
\partial_{\mu}{\left(\sqrt{-g}e^{\frac{-4\alpha\Phi}{3}}\partial_{Y}L(Y)F^{\mu \nu }
\right)}=0. \label{FE3}
\end{equation}
We would like to find perturbative solutions of the above field equations.
Our seed solution is the
the $5$-dimensional MP solution \cite{Myer}
restricted to the case where
two possible angular momenta have
equal magnitude.
Using coordinates $(t,r,\theta,\varphi_1,\varphi_2)$,
we then employ the
following parametrization for the metric \cite{Navarro}
\begin{eqnarray}\label{metric1}
ds^2 &=& g_{tt} dt^2+\frac{dr^2}{W} +
r^2\left(d\theta^{2}+\sin^{2}\theta d\varphi^{2}_{1}+
\cos^{2}\theta d\varphi^{2}_{2}\right)
+N\left(\varepsilon_{1}\sin^{2}\theta d\varphi_{1}+
\varepsilon_{2}\cos^{2}\theta d\varphi_{2}\right)^{2}\nonumber \\
&&-2B\left(\varepsilon_{1}\sin^{2}\theta d\varphi_{1}+
\varepsilon_{2}\cos^{2}\theta d\varphi_{2}\right)dt,
\end{eqnarray}
where $\varepsilon_{k}$ denotes the sense of
rotation in the $k$-th orthogonal plane of rotation,
such that $\varepsilon_{k}=\pm1$, $k=1,2$. An adequate
parametrization for the gauge potential is given by
\begin{eqnarray}\label{A1}
A_{\mu}dx^{\mu} &=&
a_{0}+a_{\varphi}\left(\varepsilon_{1}\sin^{2}\theta
d\varphi_{1}+\varepsilon_{2}\cos^{2}\theta d\varphi_{2}\right).
\end{eqnarray}
We further assume the metric functions $g_{tt}$, $W$, $N$, $B$, the two functions $a_{0}$, $a_{\varphi}$ for the gauge field, and
the dilaton function $\Phi$ depend only on the radial coordinate $r$.

We now consider perturbations around the MP solution, with an
electric charge $q$ as the perturbative parameter. In the presence
of the BI and dilaton fields, we obtain the perturbation series in
the general form
\begin{eqnarray}\label{g_{tt}}
g_{tt} = -1+\frac{2\hat{M}}{r^{2}}+q^{2}g^{(2)}_{tt}+
q^{4}g^{(4)}_{tt}+O(q^{6}) ,
\end{eqnarray}
\begin{eqnarray}\label{W}
W =1-\frac{2\hat{M}}{r^{2}}+
\frac{2\hat{J}^{2}}{\hat{M}r^{4}}+q^{2}W^{(2)}+q^{4}W^{(4)}+O(q^{6}) ,
\end{eqnarray}
\begin{eqnarray}\label{N}
N = \frac{2\hat{J}^{2}}{\hat{M}r^{2}}+q^{2}N^{(2)}+q^{4}N^{(4)}+O(q^{6}) ,
\end{eqnarray}
\begin{eqnarray}\label{B}
B = \frac{2\hat{J}}{r^{2}}+q^{2}B^{(2)}+q^{4}B^{(4)}+O(q^{6}) ,
\end{eqnarray}
\begin{eqnarray}\label{Phi}
\Phi = q^{2}\Phi^{(2)}+q^{4}\Phi^{(4)}+O(q^{6}) ,
\end{eqnarray}
\begin{eqnarray}\label{a0}
a_{0} = q a^{(1)}_{0}+q^{3} a^{(3)}_{0}+O(q^{5}) ,
\end{eqnarray}
\begin{eqnarray}\label{aph}
a_{\varphi} = q a^{(1)}_{\varphi}+q^{3} a^{(3)}_{\varphi}+O(q^{5}) ,
\end{eqnarray}
where $\hat{M}$ and $\hat{J}$ are the mass and angular momenta of
the extremal MP solution, respectively. We now fix the angular
momentum at any perturbative order, and impose the extremality
condition in all orders. We also assume that the horizon is
regular. With these assumptions, we are able to fix all constants
of integration. To simplify the notation, we introduce a parameter
$\nu$ through the equations
\begin{equation}
\hat{M}=2\nu^{2}\quad,\quad \hat{J}=2\nu^{3}\,,
\label{nu}
\end{equation}
such that the extremal MP
solution in five dimensions holds. Then, we obtain for the metric and the dilaton field the perturbation expansion up to 5th order in
the perturbative parameter $q$ and for the gauge potential functions up to 4th order
\begin{eqnarray}\label{gg_{tt}}
g_{tt} &=& -1+\frac{4\nu^{2}}{r^{2}}+
\frac{(r^{2}-4\nu^{2})q^{2}}{3\nu^{2}r^{4}}
+\Bigg{\{}{\frac {1}{2160}}\,{\frac {237-180\,{\nu}^{2}{\beta}^{2}{\alpha}^{2}+
660\,{\beta}^{2}{\nu}^{2}}{{\nu}^{8}{\beta}^{2}{r}^{2}}}+{\frac {11}{135}}\,{\frac {1}{{\beta}^{2}{\nu}^{2}{r
}^{8}}}\nonumber \\
&&+{\frac {1}{
2160}}\,{\frac {480\,{\nu}^{4}{\beta}^{2}{\alpha}^{2}-1920\,{\nu}^{4}{
\beta}^{2}-720\,{\nu}^{2}}{{\nu}^{8}{\beta}^{2}{r}^{4}}}+{\frac {1}{
2160}}\,{\frac {960\,{\beta}^{2}{\nu}^{6}+316\,{\nu}^{4}}{{r}^{6}{\nu}
^{8}{\beta}^{2}}}+{\frac {58}{135}}\,{\frac {1}{{\beta}^{2}{r}^{10}}}\nonumber \\
&&-{\frac {16
}{45}}\,{\frac {{\nu}^{2}}{{\beta}^{2}{r}^{12}}}+\left({\frac {1}{18{\beta}^{2}{\nu}^{10}}}-{\frac {{\alpha}^{2}-4}{27{\nu}^{8}}} \right)  \left( 1-2\,{\frac {{\nu}^{2}}{{r}^{2}}}
 \right) ^{2}\ln  \left( 1-2\,{\frac {{\nu}^{2}}{{r}^{2}}} \right)
\Bigg{\}}q^{4} +O(q^{6}) ,
\end{eqnarray}
\begin{eqnarray}\label{WW}
W &=&1-\frac{4\nu^{2}}{r^{2}}+\frac{4\nu^{4}}{r^{4}}-
\frac {q^{2}(r^{2}-2\nu^{2})}{3\nu^{2}r^{4}}+
\Bigg{\{}{\frac {1}{2160}}\,{\frac {522-120\,{\nu}^{2}{\beta}^{2}{\alpha}^{2}+
480\,{\beta}^{2}{\nu}^{2}}{{\nu}^{10}{\beta}^{2}}}+{\frac {44}{45}}\,{\frac {{\nu}^{2}}{{\beta}^{2}{r}^{12}}}\nonumber \\
&&-{\frac {1}{2160}}\,
{\frac {-620\,{\nu}^{4}{\beta}^{2}{\alpha}^{2}+2420\,{\nu}^{4}{\beta}^
{2}+2607\,{\nu}^{2}}{{\nu}^{10}{\beta}^{2}{r}^{2}}}-{\frac {1}{2160}}
\,{\frac {680\,{\nu}^{6}{\beta}^{2}{\alpha}^{2}-3620\,{\nu}^{6}{\beta}
^{2}-3840\,{\nu}^{4}}{{\nu}^{10}{\beta}^{2}{r}^{4}}}\nonumber \\
&&-{\frac {1}{2160}}
\,{\frac {400\,{\nu}^{8}{\beta}^{2}{\alpha}^{2}+1280\,{\nu}^{8}{\beta}
^{2}+1032\,{\nu}^{6}}{{\nu}^{10}{\beta}^{2}{r}^{6}}}-{\frac {1}{2160}}
\,{\frac {-480\,{\beta}^{2}{\alpha}^{2}{\nu}^{10}+432\,{\nu}^{8}}{{\nu
}^{10}{r}^{8}{\beta}^{2}}}-{\frac {23}{45}}\,{\frac {1}{{\beta}^{2}{r}
^{10}}}\nonumber \\
&&-{\frac {56}{45}}\,{\frac {{\nu}^{4}}{{\beta}^{2}{r}^{14}}}+ \left( {
\frac {29}{240}}\,{\frac {1}{{r}^{4}{\beta}^{2}{\nu}^{12}}}-{
\frac {{\alpha}^{2}-4}{36{\nu}^{10}{r}^{4}}} \right)  \left( -2\,{\nu}^{
2}+{r}^{2} \right) ^{3}\ln  \left( 1-2\,{\frac {{\nu}^{2}}{{r}^{2}}}
 \right)  \Bigg{\}}q^{4}+O(q^{6}) ,
\end{eqnarray}
\begin{eqnarray}\label{NN}
N &=& \frac{4\nu^{4}}{r^{2}}-\frac{2q^{2}(r^{2}+2\nu^{2})}{3r^{4}}+
\Bigg{\{}{\frac { \left( 60\,{\nu}^{2}{\beta}^{2}{\alpha}^{2
}-240\,{\beta}^{2}{\nu}^{2}-261 \right) {r}^{2}}{1080{\nu}^{10}{\beta}^{2}
}}+\,{\frac {-60\,{\nu}^{4}{\beta}^{2}{\alpha}^{2}+
240\,{\nu}^{4}{\beta}^{2}+261\,{\nu}^{2}}{1080{\nu}^{10}{\beta}^{2}}}\nonumber \\
&&+{
\frac {1}{1080}}\,{\frac {210\,{\nu}^{6}{\beta}^{2}+864\,{\nu}^{4}}{{
\nu}^{10}{\beta}^{2}{r}^{2}}}+{\frac {1}{1080}}\,{\frac {240\,{\beta}^
{2}{\nu}^{8}{\alpha}^{2}-480\,{\beta}^{2}{\nu}^{8}-732\,{\nu}^{6}}{{
\nu}^{10}{\beta}^{2}{r}^{4}}}+{\frac {1}{1080}}\,{\frac {480\,{\nu}^{
10}{\beta}^{2}+24\,{\nu}^{8}}{{\nu}^{10}{r}^{6}{\beta}^{2}}}\nonumber \\
&&+{\frac {4
}{135}}\,{\frac {1}{{\beta}^{2}{r}^{8}}}+{\frac {52}{135}}\,{\frac {{
\nu}^{2}}{{r}^{10}{\beta}^{2}}}-{\frac {16}{45}}\,{\frac {{\nu}^{4}}{{
\beta}^{2}{r}^{12}}}
\Bigg{\}}q^{4}+O(q^{6}) ,
\end{eqnarray}
\begin{eqnarray}\label{BB}
B &=& \frac{4\nu^{3}}{r^{2}}-
\frac{4\nu}{3r^{4}}q^{2}+
\Bigg{\{}{\frac {1}{540}}\,{\frac {15\,{\nu}^{2}{\beta}^{2}{\alpha}^{2}-60\,{
\beta}^{2}{\nu}^{2}-108}{{\nu}^{9}{\beta}^{2}}}+{\frac {1}{540}}\,{
\frac {-45\,{\nu}^{4}{\beta}^{2}{\alpha}^{2}+180\,{\nu}^{4}{\beta}^{2}
+324\,{\nu}^{2}}{{\nu}^{9}{\beta}^{2}{r}^{2}}}\nonumber \\
&&+{\frac {1}{540}}\,{
\frac {-360\,{\nu}^{6}{\beta}^{2}+120\,{\nu}^{6}{\beta}^{2}{\alpha}^{2
}-261\,{\nu}^{4}}{{\nu}^{9}{\beta}^{2}{r}^{4}}}+{\frac {1}{540}}\,{
\frac {240\,{\beta}^{2}{\nu}^{8}+54\,{\nu}^{6}}{{\nu}^{9}{\beta}^{2}{r
}^{6}}}+{\frac {2}{27}}\,{\frac {1}{{\beta}^{2}\nu\,{r}^{8}}}+{\frac {
52}{135}}\,{\frac {\nu}{{r}^{10}{\beta}^{2}}}\nonumber \\
&&-{\frac {16}{45}}\,{
\frac {{\nu}^{3}}{{\beta}^{2}{r}^{12}}}+( {\frac {1}{1080}}\,{\frac {15\,{\nu}^{2}{\beta}^{2}{\alpha}^{2}-60\,{\beta}^{2}{\nu}^{2}-108}{{\beta}^{2}{\nu}^{11}}}+{\frac {1}{
1080}}\,{\frac {120\,{\nu}^{4}{\beta}^{2}-30\,{\nu}^{4}{\beta}^{2}{
\alpha}^{2}+216\,{\nu}^{2}}{{\beta}^{2}{\nu}^{11}{r}^{2}}}\nonumber \\
&&+{\frac {1}{
1080}}\,{\frac {-320\,{\nu}^{6}{\beta}^{2}+80\,{\nu}^{6}{\beta}^{2}{
\alpha}^{2}-120\,{\nu}^{4}}{{\nu}^{11}{\beta}^{2}{r}^{4}}})
 \left( -2\,{\nu}^{2}+{r}^{2} \right) \ln  \left( 1-2\,{\frac {{\nu}^{
2}}{{r}^{2}}} \right)
\Bigg{\}}q^{4}+O(q^{6}) ,
\end{eqnarray}
\begin{eqnarray}\label{Phi00}
\Phi &=&-{\frac {\alpha}{4{r}^{2}{\nu}^{2}}}q^{2}+\Bigg{\{}{\frac {1}{2880}}\,{\frac {\alpha\, \left( -117+120\,{\beta}^{2}{\nu}^
{2}{\alpha}^{2}-360\,{\beta}^{2}{\nu}^{2} \right) }{{\nu}^{8}{\beta}^{
2}{r}^{2}}}\nonumber \\
&&+{\frac {1}{2880}}\,{\frac {\alpha\, \left( 126\,{\nu}^{2}+
120\,{\nu}^{4}{\beta}^{2} \right) }{{\nu}^{8}{\beta}^{2}{r}^{4}}}+{
\frac {5}{144}}\,{\frac {\alpha}{{r}^{6}{\nu}^{4}{\beta}^{2}}}+{\frac
{17}{360}}\,{\frac {\alpha}{{r}^{8}{\beta}^{2}{\nu}^{2}}}+{
\frac {\alpha}{30{\beta}^{2}{r}^{10}}}\nonumber \\
&&+{\frac {1}{144}}\,{\frac {\alpha\, \left( -3+2\,{\beta}^{2}{\nu}^{2}{
\alpha}^{2}-8\,{\beta}^{2}{\nu}^{2} \right)  \left( -2\,{\nu}^{2}+{r}^
{2} \right) }{{r}^{2}{\nu}^{10}{\beta}^{2}}}
\ln  \left( 1-2\,{\frac {{\nu}^{2}}{{r}^{2}}} \right)\Bigg{\}}q^{4}+O(q^{6})  ,
\end{eqnarray}
\begin{eqnarray}\label{a00}
a_{0} &=& \frac{q}{r^{2}}+\Bigg{\{}-{\frac {1}{180}}\,{\frac {-15+10\,{\beta}^{2}{\nu}^{2}{\alpha}^{2}-40
\,{\beta}^{2}{\nu}^{2}}{{\nu}^{6}{\beta}^{2}{r}^{2}}}-{\frac {1}{180}}
\,{\frac {40\,{\nu}^{4}{\beta}^{2}+20\,{\nu}^{4}{\beta}^{2}{\alpha}^{2
}+15\,{\nu}^{2}}{{\nu}^{6}{\beta}^{2}{r}^{4}}}-{\frac {11}{180}}\,{
\frac {1}{{\beta}^{2}{\nu}^{2}{r}^{6}}}\nonumber \\
&&-{\frac {26}{45}}\,{\frac {1}{{
\beta}^{2}{r}^{8}}}+{\frac {2{\nu}^{2}}{3{\beta}^{2}{r}^{10}}}
+{\frac { \left( 3-2\,{\beta}^{2}{\nu}^{2}{\alpha}^{2}
+8\,{\beta}^{2}{\nu}^{2} \right)  \left( -2\,{\nu}^{2}+{r}^{2}
 \right) }{72{r}^{2}{\nu}^{8}{\beta}^{2}}}\ln  \left( 1-2\,{\frac {{\nu}^{2}}{{r}^{2}}} \right)
\Bigg{\}}q^{3}+O(q^{5})  ,
\end{eqnarray}
\begin{eqnarray}\label{aphiphi}
a_{\varphi}&=&-\frac{\nu q}{r^{2}}+\Bigg{\{}{\frac {1}{720}}\,{\frac {-30+20\,{\beta}^{2}{\nu}^{2}{\alpha}^{2}-80
\,{\beta}^{2}{\nu}^{2}}{{\nu}^{7}{\beta}^{2}}}+{\frac {1}{720}}\,{
\frac {40\,{\nu}^{4}{\beta}^{2}{\alpha}^{2}-40\,{\nu}^{4}{\beta}^{2}-
27\,{\nu}^{2}}{{\nu}^{7}{\beta}^{2}{r}^{2}}}\nonumber \\
&&+{\frac {1}{720}}\,{\frac
{160\,{\nu}^{6}{\beta}^{2}+88\,{\nu}^{4}+80\,{\nu}^{6}{\beta}^{2}{
\alpha}^{2}}{{\nu}^{7}{\beta}^{2}{r}^{4}}}+{\frac {7}{60}}\,{\frac {1}
{\nu\,{\beta}^{2}{r}^{6}}}+{\frac {26}{45}}\,{\frac {\nu}{{\beta}^{2}{
r}^{8}}}-{\frac {2{\nu}^{3}}{3{\beta}^{2}{r}^{10}}}\nonumber \\
&&+{\frac {1}{144}}\,{\frac { \left( -3+2\,{\beta}^{2}{\nu}^{2}{\alpha}^{
2}-8\,{\beta}^{2}{\nu}^{2} \right)  \left( {r}^{4}-4\,{\nu}^{4}
 \right) }{{r}^{2}{\beta}^{2}{\nu}^{9}}}\ln  \left( 1-2\,{\frac {{\nu}^{2}}{{r}^{2}}} \right)\Bigg{\}}q^{3}+O(q^{5})
\end{eqnarray}
One may note that in Maxwell-dilaton's limit, $\beta
\longrightarrow \infty $, these perturbative solutions reduce to
the five dimensional perturbative charged rotating black holes in
Einstein-Maxwell-dilaton theory \cite{Allahverdi2}. A consistent
check of this solution can be provided by Smarr's formula.
%%%%%%%%%%%%%%%%%%%%%%%%%%%%%%%%%%%%%%%%%%%%%%%%%%%%%%%%%%%%%%%%%%%%%%%%
\section{Physical Quantities\label{PQ}}
In order to interpret our solutions as black holes, we should look for the event horizon. The event horizon of our rotating Born-Infeld-dilaton black holes is located at
\begin{eqnarray}\label{rh}
r_{H} &=&\sqrt{2\nu}+\frac {\sqrt{2}{q}^{2} }{24{\nu}^{3}}+\frac {{q}^{4}\sqrt{2}\left(11-8\alpha^{2}\right) }{1152{\nu}^{7}}+O(q^{6}),
\end{eqnarray}
One can see that the horizon radius does not depend on $\beta$ at least up to fourth order in the charge, while the dilaton coupling constant $\alpha$  modifies the horizon radius. In the absence of a nontrivial dilaton $ \alpha=0$, the horizon radius takes the value $r_{H} =\sqrt{2\nu}+\frac {\sqrt{2}{q}^{2} }{24{\nu}^{3}}+\frac {{q}^{4}\sqrt{2}11 }{1152{\nu}^{7}}+O(q^{6})$, exactly the result obtained for the five-dimensional perturbative Einstein-Maxwell-Born-Infeld black hole \cite{Allahverdi3}. As shown in Fig.~1, [see also Eq. (27)] one can see for large $\alpha$ the horizon radius turns negative which is unacceptable and physically meaningless. We analyze this result below, when we comment on the mass. The mass $M$, the angular momenta $J$, the dilaton charge $\Sigma$, the electric charge $Q$, and the
magnetic moment $\mu_{\rm mag}$ can be read off the asymptotic
behavior of the metric and the gauge potential \cite{Kunz2}.
The asymptotic forms are,
\begin{eqnarray}\label{quantities}
g_{tt}=-1+\frac{\tilde{M}}{r^{2}}+...,
\quad B=\frac{\tilde{J}}{r^{2}}+...,
\quad \Phi=\frac{\tilde{\Sigma}}{r^{2}}+...,
 \quad  a_{0}=\frac{\tilde{Q}}{r^{2}}+...,  \quad
a_{\varphi}=\frac{\tilde{\mu}_{\rm mag}}{r^{2}}+...,
\end{eqnarray}
where
\begin{eqnarray}\label{quantities1}
\tilde{M}=\frac{16\pi G_{5}}{3A}M,  \quad \tilde{J}=\frac{8\pi
G_{5}}{A}J, \quad \tilde{\Sigma}=\frac{4\pi G_{5}}{2A}\Sigma, \quad \tilde{Q}=\frac{4\pi G_{5}}{2A}Q,
\quad \tilde{\mu}_{\rm mag}=\frac{4\pi G_{5}}{2A}\mu_{\rm mag},
\end{eqnarray}
and $A$ is the area of the unit $3$-sphere. Comparing the above
expansions to the asymptotic behavior of the solutions, Eqs.
(\ref{gg_{tt}})-(\ref{aphiphi}), we obtain
\begin{eqnarray}\label{mass}
M =\frac{3\pi\nu^{2}}{2}+\frac{\pi q^{2}}{8\nu^{2}}+
\frac{\pi q^{4}
\left(20\nu^{2}\beta^{2}-20\nu^{2}\beta^{2}\alpha^{2}-3\right)}{5760\beta^{2}\nu^{8}}+O(q^{6}) ,
\end{eqnarray}
\begin{eqnarray}\label{charge}
Q =\pi q ,
\end{eqnarray}
\begin{eqnarray}\label{ang-mom}
J =\pi \nu^{3} ,
\end{eqnarray}
\begin{eqnarray}\label{dil-charge}
\Sigma =-{\frac {\pi \,\alpha\,{q}^{2}}{4{\nu}^{2}}}+{\frac {1}{2880}}\,{
\frac {\pi \,\alpha\,{q}^{4} \left( 40\,{\beta}^{2}{\alpha}^{2}{\nu}^{
2}-40\,{\nu}^{2}{\beta}^{2}+3 \right) }{{\beta}^{2}{\nu}^{8}}}+O(q^{6}),
\end{eqnarray}
\begin{eqnarray}\label{mag-mom}
\mu_{\rm mag} = \pi \nu q-\frac{\pi
q^{3}\left(40\nu^{2}\beta^{2}+20\nu^{2}\beta^{2}\alpha^2+3\right)}{720\nu^{5}\beta^{2}}+O(q^{5}),
\end{eqnarray}
The gyromagnetic ratio $g$ is then given by
\begin{eqnarray}\label{g}
g &=&\frac{2M\mu_{\rm mag}}{QJ}\nonumber \\
&&=3-{\frac {{q}^{2} \left( 20\,{\beta}^{2}{\alpha}^{2}
{\nu}^{2}-20\,{\nu}^{2}{\beta}^{2}+3 \right) }{240{\beta}^{2}{\nu}^{6}}}-
{\frac {{q}^{4} \left( 20\,{\beta}^{2}{\alpha}^{2}{
\nu}^{2}+10\,{\nu}^{2}{\beta}^{2}+3 \right) }{1440{\nu}^{10}{\beta}^{2}}}
+O(q^{6}),
\end{eqnarray}

It was shown that the dilaton \cite{Allahverdi2} modifies the
gyromagnetic ratio of the perturbative solutions of asymptotically
flat charged rotating dilaton black holes. In the absence of
dilaton field, it was also shown that the BI parameter  modifies
the gyromagnetic ratio of these perturbative solutions
\cite{Allahverdi3}. Here, we study the effect of the dilaton
coupling as well as the BI parameters on the gyromagnetic ratio of
our solutions simultaneously.

It is shown the behavior of the mass $M$, the magnetic moment
$\mu_{\rm mag}$ and the gyromagnetic ratio $g$ of these EBId black
holes versus $\beta$ in Fig.~2, Fig.~3  and Fig.~4 respectively.
From these figures we find out that the mass $M$, the magnetic
moment $\mu_{\rm mag}$ and the gyromagnetic ratio $g$ increase
with increasing $\beta$. We see that for some low value of $\beta$
the gyromagnetic ratio and magnetic moment $\mu_{\rm mag}$ is
zero, and then turn negative. One can speculate that this change
of sign comes from the differences of the charge distribution for
different value of the Born-Infeld parameter $\beta$ (see
\cite{Allahverdi3}). In addition these figures show that for a fixed value of $\beta$, the mass, magnetic moment and gyromagnetic ratio increase with decreasing $\alpha$. This is in agreement
with the arguments in \cite{Allahverdi2}. In the absence of a
nontrivial dilaton $\alpha=0$, the mass, the magnetic moment and
the gyromagnetic ratio reduce to
\begin{eqnarray}\label{g_navarro}
&&M =\frac{3}{2}\pi \,{\nu}^{2}+{\frac {\pi \,{q}^{2}}{8{\nu}^{2}}}-{\frac {1
}{5760}}\,{\frac {\pi{q}^{4} \, \left( -20\,{\nu}^{2}{\beta}^{2}+3\,\right) }{{\beta}^{2}{\nu}^{8}}}+O(q^{6}),\\
&&\mu_{\rm mag} =\pi \,\nu\,q-{\frac {1}{720}}\,{\frac {\pi \,q^{3} \left( 40\,{\nu}
^{2}{\beta}^{2}+3\, \right) }{{\nu}^{5}{\beta}^{2}}}+O(q^{5}),\\
 &&g =3-{\frac {1}{240}}\,{\frac {{q}^{2} \left( 3-20\,{\nu}^{2}{\beta}^{2}
 \right) }{{\beta}^{2}{\nu}^{6}}}-{\frac {1}{1440}}\,{\frac {{q}^{4}
 \left( 3+10\,{\nu}^{2}{\beta}^{2} \right) }{{\nu}^{10}{\beta}^{2}}}+O(q^{6}),
\end{eqnarray}
which are exactly the result obtained for the five-dimensional
perturbative Einstein BI black hole \cite{Allahverdi3}.
\begin{figure}[tbp]
\epsfxsize=7cm \centerline{\epsffile{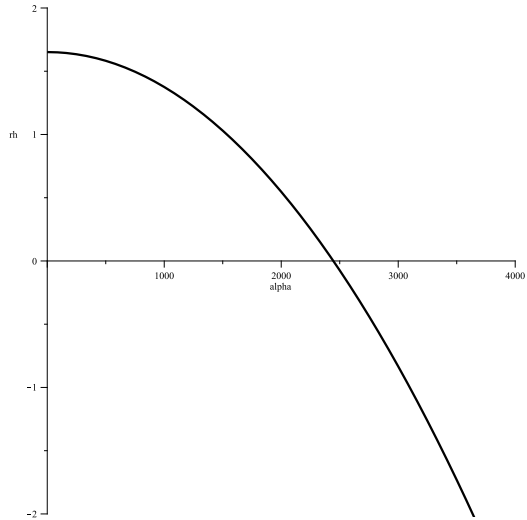}} \caption{The horizon radius
$r_{H}$ versus dilaton coupling constant $\alpha$ for $\nu=1.16$ and $q=0.09$.} \label{fig1}
\end{figure}
\begin{figure}[tbp]
\epsfxsize=7cm \centerline{\epsffile{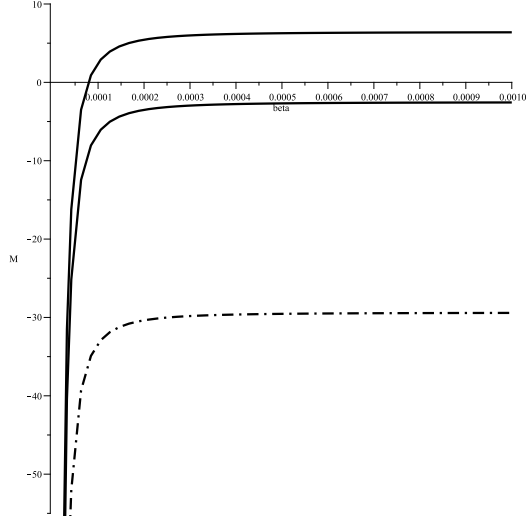}} \caption{The mass
$M$ versus BI parameter $\beta$ for $\nu=1.16$ and $q=0.09$.
$\alpha=1$ (bold line), $\alpha=5000$ (continuous line), and
$\alpha=10000$ (dotted line).} \label{fig2}
\end{figure}
\begin{figure}[tbp]
\epsfxsize=7cm \centerline{\epsffile{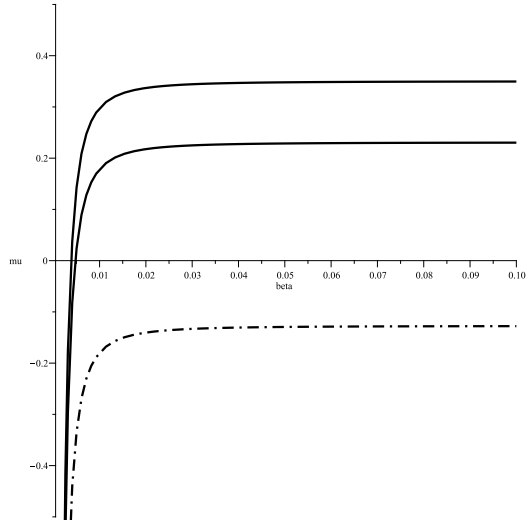}} \caption{The
magnetic moment $\mu_{\rm mag}$ versus $\beta$ for $\nu=1.16$ and
$q=0.09$. $\alpha=1$ (bold line), $\alpha=50$ (continuous
line), and $\alpha=100$ (dotted line).} \label{fig3}
\end{figure}
\begin{figure}[tbp]
\epsfxsize=7cm \centerline{\epsffile{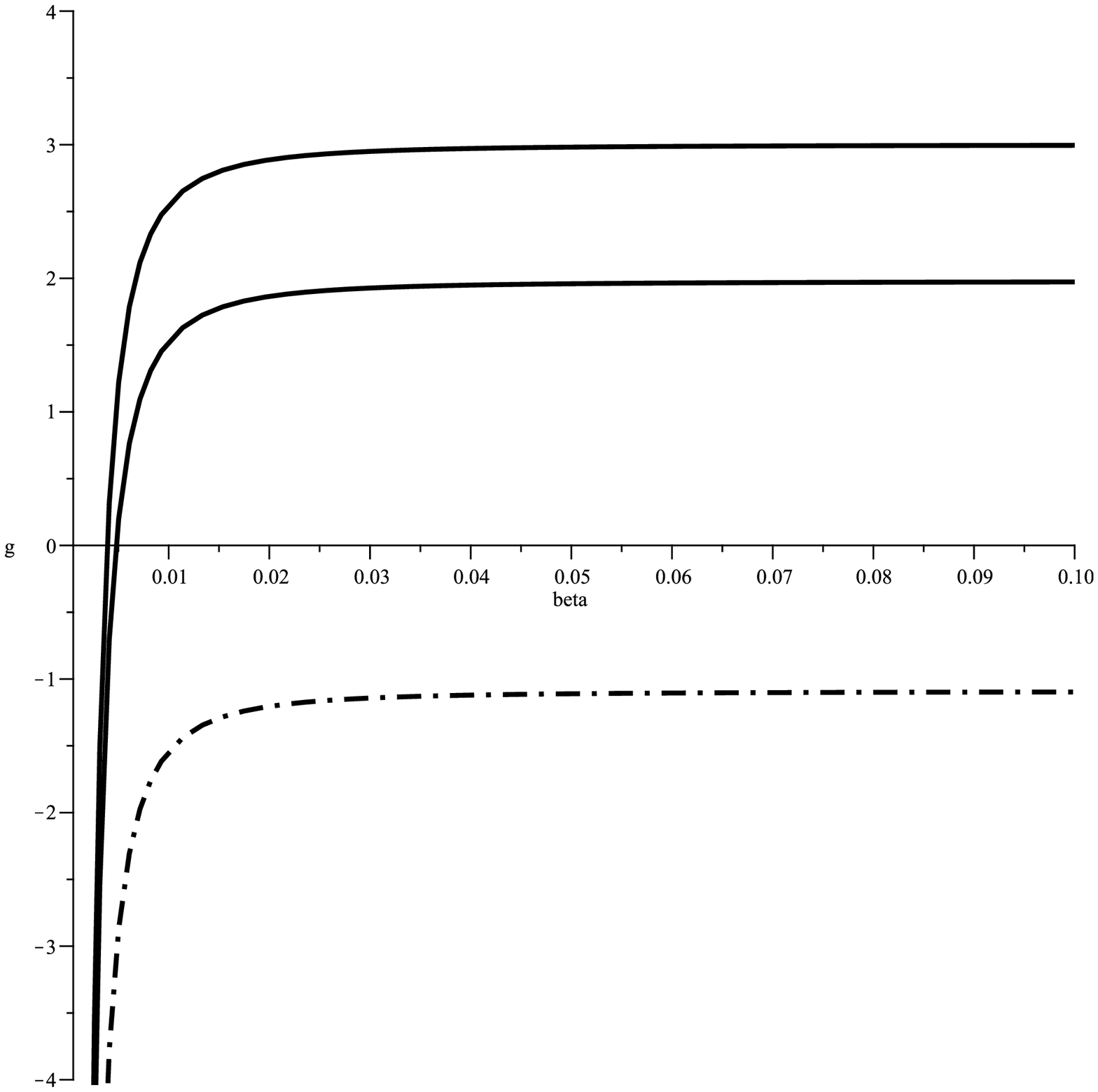}} \caption{The
gyromagnetic ratio $g$ versus BI parameter $\beta$ for $\nu=1.16$
and $q=0.09$. $\alpha=1$ (bold line), $\alpha=50$ (continuous
line), and $\alpha=100$ (dotted line).} \label{fig4}
\end{figure}
The mass $M$ of the black holes as a function of the Born-Infeld parameter $\beta$ and the dilaton coupling constant $\alpha$ exhibits interesting behavior, as shown in Fig. 2, [see also Eq. (30)].  For large $\alpha$ the mass is negative. We do not attach any significance to this result since the value of $\alpha$ for which the mass is negative not acceptable as discussed before. On the other hand, for very small $\beta$ the mass turns negative. We also do not attach any significance to this result since the value of $\beta$ for which the mass is zero uses a perturbative  $q^{2}$  term that is much larger than the zeroth-order term in the expressions for the magnetic moment and gyromagnetic ratio. In fact, the values of  $\beta$ for which the results make thorough sense are values of  $\beta$ larger than the ones which yield the zeros of the magnetic moment and gyromagnetic ratio.
Define $\xi$ as the timelike Killing vector and $\eta_{k}$,
$k=1,2$, as the two azimuthal Killing vectors. The two equal
horizon constant angular velocities $\Omega$ can then be defined
by imposing that the Killing vector field
\begin{eqnarray}\label{chi}
\chi = \xi+\Omega\sum^{2}_{k=1}\epsilon_{k}\eta_{k},
\end{eqnarray}
is null on the horizon and orthogonal to it as well.
This yields,
\begin{eqnarray}\label{Omega}
\Omega = \frac{1}{2\nu}-\frac{q^2}{24\nu^{5}}-
\frac{q^{4}
\left(5\nu^{2}\beta^{2}-5\nu^{2}\beta^{2}\alpha^2-1\right)}{1440\beta^{2}\nu^{11}}+O(q^{6})\,.
\end{eqnarray}
The area of the horizon $A_{H}$ and the
electrostatic potential at the horizon $\Phi_{H}$ are given by
\begin{eqnarray}\label{AH}
A_{H} =8\pi^{2}\nu^{3}+O(q^{6})\, .
\end{eqnarray}
\begin{eqnarray}\label{PhiH}
\Phi_{H} ={\frac {q}{4{\nu}^{2}}}-{\frac {1}{1440}}\,{\frac {{q}^{3} \left(
20\,{\beta}^{2}{\alpha}^{2}{\nu}^{2}-20\,{\nu}^{2}{\beta}^{2}+3
 \right) }{{\beta}^{2}{\nu}^{8}}}+O(q^{5})\,.
\end{eqnarray}
The surface gravity $\kappa$ is  defined by $
\kappa^{2}=-\frac{1}{2}(\nabla_{\mu}
\chi_{\nu})(\nabla^{\mu}\chi^{\nu}) $. Taking into account the
conserved quantities obtained in this section, one can check that
these quantities satisfy the Smarr mass formula up to 4th order
\begin{eqnarray}\label{smarr}
M = 3\Omega
J+2\Phi_{H}Q+\frac{\beta }{2}\frac{\partial M}{\partial \beta}+\frac{\Sigma}{\alpha},
\end{eqnarray}
For four dimensional Born-Infeld black holes it has been defined a
new thermodynamic quantity $\frac{\partial M}{\partial \beta}$
conjugate to the Born-Infeld parameter $\beta$, this quantity is
required for consistency of the Smarr relation \cite{Gunasekaran}.
Recently, we have generalized this results for perturbative
Einstein-Maxwell-Born-Infeld black holes in five dimensions
\cite{Allahverdi3}. Here, we see the effect of the dilaton as well
as Born-Infeld parameter in the Smarr relation simultaneously
Eq.(\ref{smarr}).
\section{Summary and Conclusions\label{SumCon}}
To sum up, we have presented a new class of perturbative charged
rotating dilaton black hole solutions in five dimensions in the
presence of nonlinear BI gauge field. Our investigations were
restricted to the extremal black holes with equal angular momenta.
We have used the perturbative method and solved the equations of
motion up to the $4$th order of the perturbative charge parameter
$q$. We have started from the rotating MP \cite{Myer} black hole
solutions. Then, we have considered the effect of adding an amount
of perturbative charge parameter $q$ to the solutions in the
presence of the dilaton and BI fields. We computed the physical
quantities of the solutions such as the mass, the angular
momentum, the dilaton charge, the electric charge and the magnetic
moment. We have also obtained the gyromagnetic ratio of these
rotating black holes. Interestingly enough, we found that the
perturbative parameter $q$, the dilaton coupling constant
$\alpha$, and the BI parameter $\beta$, modify the values of the
physical quantities. One can see that for a certain value of the
dilaton coupling $\alpha$, for small $\beta$ the mass $M$ is
negative and it increases with increasing $\beta$ while, in the
linear Maxwell field regime it is always positive. We know that there is a critical value $\beta_{c}$ in which for small values of $\beta < \beta_{c}$, a non-extreme horizon cover the black hole. (see \cite{RN-Hendi} for more details). In addition, we
found out that for constant $\alpha$, the gyromagnetic ratio is
negative for small $\beta$ and increases with increasing $\beta$
while, for large $\beta$ it tends to a constant value. We computed
the conserved quantities of the solutions and verified that these
quantities satisfy the Smarr mass formula of the black holes.

Finally, we would like to mention that in this paper, we only
studied the  extremal perturbative charged rotating EBId black
holes in five dimensions. The generalization of the present work
to all higher dimensional case is now under investigation and will
be addressed elsewhere.
%%%%%%%%%%%%%%%%%%%%%%%%%%%%%%%%%%%%%%%%%%%%%%%%%%%%%%%%%%%%%%%%%%%%%%%
\acknowledgments{  M. Allaverdizadeh is supported by a FCT grant.  S.H.H and A.S. thank Shiraz University Research Council. The works
 of S.H.H and A.S. have been supported financially by Center for
Excellence in Astronomy and Astrophysics of IRAN (CEAAI-RIAAM).
%%%%%%%%%%%%%%%%%%%%%%%%%%%%%%%%%%%%%%%%%%%%%%%%%%

\end{document}